\documentstyle[aps,prd,eqsecnum,preprint,tighten]{revtex}
\begin{document}
\draft
\preprint{OKHEP-97-04}
\title{Observability of the Bulk Casimir Effect:
Can the Dynamical Casimir Effect be Relevant to Sonoluminescence?}
\author{Kimball A. Milton\thanks{e-mail: milton@mail.nhn.ou.edu}}
\address{ Department of Physics and Astronomy,
The University of Oklahoma, Norman OK 73019 USA}
\author{Y. Jack Ng\thanks{e-mail: ng@physics.unc.edu}}
\address{Institute of Field Physics, Department of Physics and
Astronomy,
University of North Carolina, Chapel Hill, NC 27599 USA}
\date{\today}
\maketitle

\begin{abstract}
The experimental observation of intense light emission by
acoustically
driven, periodically collapsing bubbles of air in water
(sonoluminescence)
has yet to receive an adequate explanation.  One of the most
intriguing
ideas is that the conversion of acoustic energy into photons occurs
quantum mechanically, through a dynamical version of the Casimir
effect.
We have argued elsewhere that in the adiabatic approximation, which
should
be reliable here, Casimir or zero-point energies cannot possibly be
large
enough to be relevant. (About 10 MeV of energy is released per
collapse.)  However, there are sufficient subtleties involved that
others
have come to opposite conclusions.  In particular, it has been
suggested
that bulk energy, that is, simply the naive sum of
${1\over2}\hbar\omega$,
which is proportional to the volume, could be relevant.  We show that
this
cannot be the case, based on general principles as well as specific
calculations.  In the process we further illuminate some of the
divergence
difficulties that plague Casimir calculations, with an example
relevant to
the bag model of hadrons.
\end{abstract}

\section{Introduction}
One of the most intriguing phenomena in physics today is
sonoluminescence \cite{sono1,sono2,sonorev}. In the experiment, a
small
(radius $\sim 10^{-3}$ cm) bubble of air or other gas is injected
into water,
and subjected to an intense acoustic field (overpressure $\sim 1$
atm,
frequency $\sim 2\times 10^4$ Hz).  If the parameters are carefully
chosen, the repetitively collapsing bubble emits an intense flash of
light at minimum radius (something like a million optical photons are
emitted per flash), yet the process is sufficiently non-catastrophic
that a single bubble may continue to undergo collapse and emission
20,000 times a second for many minutes, if not months.  Many curious
properties have been observed, such as sensitivity to small
impurities,
strong temperature dependence, necessity of small amounts of noble
gases,
possible strong isotope effect, etc.

No convincing theoretical explanation of the light-emission process
has
yet been put forward.  This is certainly not for want of interesting
theoretical ideas \cite{theory}.  One of the most intriguing
suggestions
was put forward by Schwinger\cite{js}, based on a reanalysis of the
Casimir effect\cite{rederiv}.  Specifically, he proposed that the
Casimir
effect, first considered by Casimir as the force between parallel
conducting
plates due to zero-point fluctuations in the
fields\cite{casimir,exp},
be generalized to the spherical volume defined by the bubble
\cite{boyer,balian,mds}, and with the static boundary conditions
appropriately
removed.  He called this, as yet, unformulated theory the dynamical
Casimir effect.  Unfortunately, although Schwinger began the general
reformulation of the static problem in Ref. \cite{rederiv} (most of
which had
been, unbeknownst to him, given earlier\cite{kim}), he did not live
to
complete the program.  Instead, he contented himself with a rather
naive approximation of subtracting the zero-point energy
${1\over2}\sum\hbar\omega$ of the medium
from that of the vacuum, leading, for a spherical bubble of radius
$a$
in a medium with index of refraction $n$, to a Casimir energy
proportional
to the volume of the bubble:
\begin{equation}
E_{\rm bulk}={4\pi a^3\over3}\int {(d{\bf k})\over(2\pi)^3}{1\over2}k
\left(1-{1\over n}\right).
\label{bulk}
\end{equation}
Of course, this is quartically divergent.  If one puts in a suitable
ultraviolet cutoff, one can indeed obtain the needed 10 MeV per
flash.
On the other hand, one might have serious reservations about the
physical
meaning of such a divergent result.

In an earlier paper, we reconsidered the Casimir effect explanation
of
sonoluminescence \cite{milng,leipzig}.  We argued there that the
leading term
(\ref{bulk}) was to be removed by subtracting the contribution the
formalism
would give if either medium filled all space.  Doing so still left us
with a cubically divergent Casimir energy; but we argued further that
this
cubic divergence could plausibly be removed as a contribution to the
surface energy.  The remaining finite energy, in the presumably
accurate uniform asymptotic approximation,
\begin{equation}
E_c\sim-{(n-1)^2\over64 a},
\label{fincas}
\end{equation}
is at least ten orders of magnitude too small to be relevant to
sonoluminescence.

The reader might object at once that all this is in the static
approximation,
and the rapidly collapsing bubbles involved in sonoluminescence are
anything
but static.  However, the time scales seems favorable for a simple
adiabatic
approximation to be accurate.  Optical photons correspond to a time
scale
$\sim 10^{-15}$ s, while the flash duration is $\sim 10^{-11}$ s.
That
is, the bubble changes very little during one period of the light
emitted.
Of course, there may be processes here occurring on much smaller time
scales, so it would be highly desirable to remove this adiabatic
approximation,
which we hope to accomplish in a subsequent publication.

Eberlein \cite{eberlein} has also proposed a version of the dynamical
Casimir mechanism (perhaps more properly called the Unruh
\cite{unruh} mechanism)  which she claims can explain the observed
radiation.  We have criticized her calculation on technical grounds
\cite{milng}, but mostly on the basis of her use of ultrarelativistic
velocities.  See also Ref.\ \cite{ebercrit}. If, in fact, reasonable
numbers are used in her result, the energies involved are too small
by
18 orders of magnitude, and even if her ultrarelativistic velocities
are used, only $10^{-3}$ MeV is available.  So, qualitatively, her
results are not inconsistent with ours.

However, recently there has been a  proposal that, indeed,
the bulk energy result of Schwinger is relevant (of course, it's
correct)
\cite{carlson}.  These authors make a great issue of the subtraction
of the uniform medium contribution, implying, it would seem, that we
were unaware of what we were doing.  Since this is a serious issue
with experimental consequences, and since, admittedly, there are
subtle
issues of principle involved here, in this paper we wish to return to
this point and provide further evidence for our result
(\ref{fincas}).
In the following section we will explain more fully why this
subtraction
was made, indicate that it has a rather long history in Casimir
effect
calculations, and was in fact made by Schwinger in \cite{rederiv}
before
he abandoned that effort.  Then, in Section III, we recall the old
connection
between the Casimir effect and van der Waal forces, and show, in
fact, that
a finite energy of the same magnitude as the Casimir energy
(\ref{fincas})
can be obtained from the latter.  Finally, motivated by recent work
on regulating Casimir energies by continuing in the
number of space dimensions \cite{benmil},
we examine, in the Appendix, whether dimensional continuation can be
used
to give an unambiguously finite value for the Casimir energy for a
bubble
in a dielectric, for example.  The negative answer to the latter
question
shows that the quartic and cubic divergences found there are real.
Again,
appropriate {\em physical\/} arguments must be used to show that they
are not
{\em relevant\/} to the situation at hand.

\section{Definition of the Casimir Energy}
In \cite{milng} we derived a formula for the Casimir energy due to
electromagnetic field fluctuations in a space divided into two parts
by
a spherical surface of radius $a$.  The interior region, $r<a$, the
inside
of the bubble, has permittivity $\epsilon'$ and permeability $\mu'$,
while the exterior region, $r>a$, the outside of the bubble, has
permittivity
$\epsilon$ and permeability $\mu$.  We initially ignore dispersion.
(Although it can be included \cite{disp}, dispersion turns out not to
affect our conclusions \cite{milng}.)
We calculate vacuum expectation values of field products in terms of
Green's dyadics for the corresponding classical electrodynamics
problem:
\begin{mathletters}
\begin{eqnarray}
i\langle {\bf E(r)}{\bf E(r'})\rangle&=&\bbox{\Gamma}
({\bf r},{\bf r'}),\\
i\langle{\bf B(r}){\bf B(r'})\rangle&=&-{1\over\omega^2}\bbox{\nabla}
\times\bbox{\Gamma}({\bf r},{\bf
r}')\times\overleftarrow{\bbox{\nabla}'},
\end{eqnarray}
\end{mathletters}
where $\bbox{\Gamma}$ is the Green's dyadic for Maxwell's equations
\cite{sdm}.  The result for the Casimir energy is
\begin{equation}
E=-{1\over4\pi a}\int_{-\infty}^\infty dy
e^{iy\delta}\sum_{l=1}^\infty
(2l+1)x{d\over dx}\ln S_l,
\label{unsuben}
\end{equation}
where
\begin{eqnarray}
S_l=[s_l(x')e'_l(x)-s_l'(x')e_l(x)]^2-\xi^2[s_l(x')
e'_l(x)+s_l'(x')e_l(x)]^2,
\end{eqnarray}
with
\begin{eqnarray}
\xi={\sqrt{{\epsilon'\mu\over\epsilon\mu'}}-1\over
\sqrt{{\epsilon'\mu\over\epsilon\mu'}}+1},
\end{eqnarray}
which is expressed in terms of modified Bessel functions
\begin{mathletters}
\begin{eqnarray}
s_l(x)&=&\sqrt{x}I_{l+1/2}(x),\\
e_l(x)&=&\sqrt{x}K_{l+1/2}(x).
\end{eqnarray}
\end{mathletters}
The expression for the energy is regulated by the insertion of a
Euclidean time-splitting
parameter, $\delta=(x_4-x_4')/a$, and the variables are
\begin{equation}
x=|y|\sqrt{\epsilon\mu},\quad x'=|y|\sqrt{\epsilon'\mu'}.
\end{equation}

It is completely manifest that (\ref{unsuben}) does not have a
well-defined
limit as $\delta\to0$ ---it is quartically divergent.  Indeed, it is
easy
to show as \cite{carlson} does, that the quartically divergent term
here
corresponds precisely to the Schwinger result (\ref{bulk}) when
$\epsilon'=
\mu'=1$, $\mu=1$.  However, it is also quite clear that the
calculation
is not yet done when we have reached this point.  As we stated in
\cite{milng},
``We must remove the term which would be present if either medium
filled
all space (the same was done in the case of parallel dielectrics
\cite{sdm}).''
When we look at the latter reference, we see immediately the point.
Again to quote, this time from \cite{sdm}: ``These terms [to be
subtracted]
correspond to the electromagnetic energy required to replace medium 1
by
medium 2 in the displacement volume.  (Since this term in the energy
is
already phenomenologically described, it must be cancelled by an
appropriate
contact term.)''  What we were saying there, in the
present context, is that the term in the energy
corresponding to the boundary-condition-independent Green's function
\begin{equation}
F^{(0)}_l=ikj_l(kr_<)h_l^{(1)}(kr_>),
\end{equation}
must be removed, because it contributes (a formally infinite amount)
to
the bulk energy of the material, which is already phenomenologically
described in terms of its bulk properties.  In fact, we are not
creating
material, e.g., water, we are simply displacing it when we insert the
bubble, and force the bubble to expand and contract.  The energy per
unit element of medium is therefore not changed.
(The density of the air in the bubble of course changes greatly,
but the zero-point energy of that relatively dilute medium is
certainly
insignificant because $n\approx1$. In any case, the effect of this
density change is also included in the phenomenological description.)

Indeed, the spectacular agreement between the the Lifshitz theory of
parallel dielectrics \cite{lifshitz}, rederived in \cite{sdm}, and
the beautiful experiment of Sabisky and Anderson \cite{anderson}
seems strong vindication of this subtraction procedure.

Further evidence that we are on the right track is provided by
Schwinger
himself.  In the first National Academy article cited in
\cite{rederiv},
where he rederives the result for parallel dielectrics, he explicitly
removes volume and surface energies:
\begin{quote}
one finds contributions to $E$ that, for example, are proportional
\dots
to the volume enclosed between the slabs.  The implied constant
energy
density---independent of the separation of the slabs---violates the
normalization of the vacuum energy density to zero.  Accordingly, the
additive
constant has a piece that maintains the vacuum energy normalization.
There is also a contribution to $E$ that is proportional to [the
area],
energy associated with individual slabs.  The normalization to zero
of the energy for an isolated slab is maintained by another part of
the
additive constant.
\end{quote}
Admittedly, the situation is more clear-cut in the parallel-plate
geometry.
However, in the following paper (the last reference in
\cite{rederiv})
where Schwinger begins to set up the problem for the spherical
geometry
(but leaves the details to Harold \cite{harold}), a close reading
shows
a similar subtraction is implicit.  Unfortunately, when Schwinger
went
on to apply Casimir energy to sonoluminescence in \cite{js}, he does
not make use of the general analysis in \cite{rederiv}.  Our
interpretation
is that at that point Schwinger lost the energy or courage to
complete
the full calculation, and needing an immediate result to confront
the phenomenology, simply jumped to the unsubtracted, unregulated
result
(\ref{bulk})---see the second reference in \cite{js}.

But enough of argumentation.  Let us turn to detailed calculations
that support our contention.

\section{Derivation of Casimir Effect from van der Waals Forces}

It is familiar that the van der Waals forces between polarizable
molecules---the Casimir-Polder forces \cite{cp}---can be derived from
the Casimir forces between dielectric bodies.  We interpret this
as meaning that the Casimir effect is merely a local field form
of the action-at-a-distance summation of the forces between the
molecules
that make up the material bodies.

Let us begin with a variation of the argument given in \cite{sdm}.
Consider
a dielectric slab bounded by planes $z=0$ and $z=a$, having
dielectric
constant $\epsilon$; outside this region there is vacuum,
$\epsilon=1$.
According to the Lifshitz formula \cite{sdm,lifshitz}, the force/area
between
the surfaces is
\begin{eqnarray}
f=-\int_0^\infty {d\zeta\over2\pi}\int_0^\infty {dk^2\over2\pi}
 \kappa_3\Bigg\{
\left[\left(\kappa_3+\kappa_1\over\kappa_3-
\kappa_1\right)^2e^{2\kappa_3a}
-1\right]^{-1}
+\left[\left(\kappa'_3+\kappa'_1\over\kappa'_3-\kappa'_1\right)^2
e^{2\kappa_3a}-1\right]^{-1}\Bigg\},
\label{lifshitzforce}
\end{eqnarray}
where, in the $i$th medium (we denote the region of the slab by 3,
that
the outside regions by 1),
\begin{equation}
\kappa_i^2=k^2+\epsilon_i\zeta^2,\quad
\kappa'_i={\kappa\over\epsilon_i}.
\end{equation}
Now suppose the medium is tenuous, so that the dielectric constant
differs
only slightly from unity,
\begin{equation}
\epsilon-1\ll1.
\end{equation}
Then, with a simple change of variable,
\begin{equation}
\kappa=\zeta p,
\end{equation}
we can recast the Lifshitz formula (\ref{lifshitzforce}) into the
form
\begin{equation}
f\approx-{1\over32\pi^2}\int_0^\infty d\zeta\,\zeta^3\int_1^\infty
{dp\over p^2}[\epsilon(\zeta)-1]^2[(2p^2-1)^2+1]e^{-2\zeta pa}.
\label{weaklif}
\end{equation}
If the separation of the surfaces is large compared to the
characteristic
wavelength characterizing $\epsilon$, $a\zeta_c\gg1$, we can
disregard
the frequency dependence of the dielectric constant,
and we find
\begin{equation}
f\approx-{23(\epsilon-1)^2\over640\pi^2a^4}.
\label{longdist}
\end{equation}
For short distances, $a\zeta_c\ll1$, the approximation is
\begin{equation}
f\approx-{1\over32\pi^2}{1\over a^3}\int_0^\infty
d\zeta(\epsilon(\zeta)-1)^2.
\label{shortdist}
\end{equation}
These formulas are identical with the well-known forces
 found for the complementary geometry in \cite{sdm}.

Now we wish to derive these results from the sum of van der Waals
forces,
derivable from a potential of the form
\begin{equation}
V=-{B\over r^\gamma}.
\end{equation}
We do this by computing the energy ($N= $density of molecules)
\begin{equation}
E=-{1\over2}B N^2\int_0^a dz\int_0^a dz'\int(d{\bf r_\perp})(d{\bf
r'_\perp})
{1\over[({\bf r_\perp-r'_\perp})^2+(z-z')^2]^{\gamma/2}}.
\end{equation}
If we disregard the infinite self-interaction terms (see below), we
get
\begin{equation}
f=-{\partial\over\partial a}{E\over A}=-{2\pi B
N^2\over(2-\gamma)(3-\gamma)}
{1\over a^{\gamma-3}}.
\end{equation}
So then, upon comparison with (\ref{longdist}), we set $\gamma=7$
and in terms of the polarizability,
\begin{equation}
\alpha={\epsilon-1\over4\pi N},
\end{equation}
we find
\begin{equation}
B={23\over4\pi}\alpha^2,
\label{bee}
\end{equation}
or, equivalently, we recover the retarded dispersion potential,
\begin{equation}
V=-{23\over4\pi}{\alpha^2\over r^7},
\end{equation}
whereas for short distances we recover the London potential,
\begin{equation}
V=-{3\over\pi}{1\over r^6}\int_0^\infty d\zeta\,\alpha(\zeta)^2.
\end{equation}

Our intention is to carry out the same simple calculation for a
dielectric
sphere.  The first couple of steps are unambiguous ($\theta$ is the
angle
between $\bf r$ and $\bf r'$):
\begin{eqnarray}
E&=&-{1\over2}BN^2\int(d{\bf r})(d{\bf
r}'){1\over(r^2+r^{\prime2}-2rr'\cos\theta
)^{\gamma/2}}\nonumber\\
&=&-{4\pi^2BN^2\over2-\gamma}\int_0^a dr\int_0^a dr'\,rr'
\left[{1\over(r+r')^{\gamma-2}}-{1\over|r-r'|^{\gamma-2}}\right].
\end{eqnarray}
Now, however, there are divergences of two types, ``volume'' ($r'\to
r$) and
``surface'' ($r\to a$).  The former is of a universal character.
If we regulate it by a naive point separation, $r'\to r+\delta$,
$\delta\to
0$, we find the most divergent part to be
\begin{equation}
E_{\rm vol}=-{\pi B N^2\over10}{1\over\delta^4}V,\quad V={4\pi
a^3\over3},
\end{equation}
which is identical to the corresponding (omitted) divergent term in
the parallel dielectric calculation, where $V=aA$.  This is obviously
the self-energy divergence that would be present if the medium filled
all space, and makes no reference to the interface, and is therefore
quite unobservable.  This is the analogue (although the $\epsilon$
dependence is different) of the volume divergence in the Casimir
effect,
(\ref{bulk}).

If, once again, the divergent terms are simply omitted, as may be
weakly
justified by continuing in the exponent $\gamma$ from $\gamma<3$, we
obtain
a positive energy,
\begin{equation}
E_{\rm vdW}={23\over1536\pi a}(\epsilon-1)^2.
\label{finvdw}
\end{equation}
This may be more rigorously justified by continuing in dimension, a
procedure which has proved useful and illuminating in Casimir
calculations
\cite{benmil}.  Thus we replace the previous expression for the
energy
by
\begin{equation}
E=-{1\over2}BN^2\int d^Dr\,d^Dr'{1\over|{\bf r-r'}|^\gamma}
\end{equation}
where, in terms of the last angle in $D$-dimensional polar
coordinates,
\begin{equation}
\int d^Dr={2\pi^{(D-1)/2}\over\Gamma\left(D-1\over2\right)}\int_0^a
dr\,r^{D-1}\int_0^\pi d\theta\sin^{D-2}\theta.
\end{equation}
If we take, say, $\bf r'$ to lie along the $z$ axis, so that $\theta$
is again
the angle between $\bf r$ and $\bf r'$, we find
\begin{eqnarray}
E&=&-{1\over2}BN^2{2\pi^{D/2}\over\Gamma\left(D\over2\right)}
{2\pi^{(D-1)/2}\over\Gamma\left(D-1\over2\right)}
\int_0^a dr'\,r^{\prime D-1}\int_0^a dr\,r^{D-1}\nonumber\\
&&\quad\times\int_{-1}^1d\cos\theta\,
(1-\cos^2\theta)^{(D-3)/2}(r^2+r^{\prime2}-
2rr'\cos\theta)^{-\gamma/2}.
\end{eqnarray}
The angular integration can be given in terms of an associated
Legrendre
function $P^a_b(z)$,
\begin{eqnarray}
&&\int_{-1}^1dt\,(1-t^2)^{(D-3)/2}(r^2+r^{\prime2}-2rr't)^{-\gamma/2}
\nonumber\\
&&\qquad=\sqrt{\pi}\Gamma\left(D-1\over2\right)
(rr')^{1-D/2}|r^2-r^{\prime2}
|^{(D-\gamma-2)/2}P_{(\gamma-D)/2}^{1-D/2}\left(r^2+r^{\prime2}\over
|r^2-r^{\prime2}|\right).
\end{eqnarray}
Now let us substitute this into the expression for the energy, and
change variables from $r$, $r'$ to
\begin{equation}
x=r^2+r^{\prime2},\quad y={r^2+r^{\prime2}\over|r^2-r^{\prime2}|}.
\end{equation}
The $x$ integral is then trivially done, leaving us with
\begin{equation}
E=-{BN^2\pi^D\over2^{D/2}\Gamma(D/2)}{1\over D-\gamma/2}
\int_1^\infty dy \left(2a^2\over
y+1\right)^{D-\gamma/2}(y^2-1)^{(D-2)/4}
P_{(\gamma-D)/2}^{1-D/2}(y),
\end{equation}
valid for $D>\gamma/2$.
Integrals of this type are given in \cite{prud}:
\begin{equation}
\int_1^\infty
dy\,(y-1)^{-a/2}(y+1)^{b+a/2-1}P_b^a(y)=2^{b}{\Gamma(-2b)\over
\Gamma(1-b-a)\Gamma(1-b)},
\end{equation}
valid for $\mbox{Re }a<1$, $\mbox{Re }b<0$.
Then we have, using the duplication formula for the $\Gamma$
function,
\begin{equation}
E=-BN^2{\pi^{D-1/2}2^{D-\gamma}\Gamma\left(D-\gamma+1\over2\right)
\over\Gamma(D/2)\Gamma(D-\gamma/2+1)(D-\gamma)}.
\label{analen}
\end{equation}
The resulting formula is regular when $D$ and $\gamma$ are both odd
integers, so we can analytically
continue from $D>\gamma$ to $D=3$ for $\gamma=7$. Doing so gives us,
using Eq.~(\ref{bee}),
\begin{equation}
E=BN^2{\pi^2\over24}{1\over a}={23\over24}{(\epsilon-1)^2\over64\pi
a},
\label{finvdw2}
\end{equation}
exactly the same as (\ref{finvdw}).  Note that the magnitude of this
result is nearly the same as that found in \cite{milng},
and stated in (\ref{fincas}),  differing
only by the factor
\begin{equation}
{23\over24}{4\over\pi}=1.22,
\end{equation}
which is a plausible difference in that the previous calculation was
only in the leading asymptotic approximation, but the sign is
opposite!
We offer as evidence for the validity of this methodology the
fact that the formula (\ref{analen}) gives the correct Coulomb
energy of a uniform ball of charge, for which $\gamma=1$.

Evidently, we have reached the frontier of our understanding of the
Casimir
effect and its connection with van der Waals forces.  The subtraction
procedure may well be ambiguous, although the volume and surface
divergences are unambiguous.  That these divergences are real is
further
reinforced by the considerations of the Appendix, which shows that
the technique
of dimensional continuation fails for this case.  But these
divergences
are not {\em relevant\/} to the light emission process, although they
would
be to a first-principles calculation of the energy density and
surface
tension of the medium \cite{sdm}.  However, our qualitative
conclusion,
that quantum vacuum energies are completely irrelevant to
sonoluminescence,
is dependent only on the order of magnitude of the finite remainder,
given by either (\ref{fincas}) or (\ref{finvdw}).

\section{Conclusions}
Our conclusions here are threefold:
\begin{itemize}
\item The divergences that occur when interior and exterior modes
are mismatched, whether by exclusion of one set, or by changing the
the speed of light in the two media, are real, and cannot be
circumvented
by a mathematical trick.
\item Volume divergences are not physically meaningful, since they
reflect self-energy effects, and serve to define the intrinsic
properties
of the material.  They are naturally cancelled out by the
introduction
of a suitable contact term.  What is left is a surface divergence,
which presumably is physically meaningful, yet should be absorbed
into
a renormalization of physical parameters, such as the surface
tension.
\item The magnitude of the finite remainder, of order $1/a$,
apparently may be extracted unambiguously.
Whatever its sign, it is far too small to be relevant to
sonoluminescence.
\end{itemize}

\section*{Acknowledgements}
We thank M. Visser for discussion about his work and the significance
of the bulk energy contribution.  We thank the U.S. Department of
Energy
for financial support of this work.

\appendix

\section*{Dimensional continuation of the Casimir effect}

The fact that the above dimensional regularization of the van der
Waals
energy gave a finite result suggests that we re-examine the Casimir
calculation
to see if possibly an unambiguous finite result could thereby be
obtained.
We will not be surprised to find a negative answer to this question,
since the perfect
cancellation between interior and exterior modes cannot hold true
with
different speeds of the light in the two media \cite{brevik}.

We will content ourselves by examining the extreme case of
$\epsilon\to\infty$
in the exterior region, that is, a bag with perfectly conducting
boundary
conditions on the surface.  Since it is necessary to continue the
individual
modes, we will examine the TE mode as representative.
[As we will see, the subleading divergences cancel between the TE and
TM
modes.]
 In three dimensions the interior modes alone give \cite{bag}
\begin{equation}
E^{\rm TE}_{\rm in}=-{1\over2\pi
a}\sum_{n=1}^\infty(2n+1)\int_0^\infty
dx\,x{s'_n(x)\over s_n(x)},
\end{equation}
where the generalized modified Ricatti-Bessel functions are
\begin{equation}
s_n(x)=x^{D/2-1}I_\nu(x),\quad e_n(x)=x^{D/2-1}K_\nu(x),
\end{equation}
where $\nu=n+D/2-1$ ($=n+1/2$ here).
The generalization of this result to $D$ space dimensions is
\cite{benmil}
\begin{equation}
E^{\rm TE}_{\rm in}=-{1\over2\pi
a}{1\over\Gamma(D-1)}\sum_{n=1}^\infty
w(n,D)\int_0^\infty dx\,x{s_n'(x)\over s_n(x)},
\end{equation}
where the weight function is
\begin{equation}
w(n,D)={2\nu\Gamma(n+D-2)\over n!}.
\end{equation}

Again, as elsewhere, in \cite{bag} the ``vacuum energy'' term was
subtracted.
(As noted in the text,
the justification was only partly that it removed the most divergent
terms.)
This was obtained from the free Green's function, which in $D$ space
dimensions
is
\begin{equation}
{\cal
G}^0_\omega(r,r',\theta)=i\sum_n{2\nu\Gamma\left({D\over2}-1\right)
\over8(\pi
rr')^{D/2-1}}C_n^{(D/2-1)}(\cos\theta)J_\nu(kr_<)H_\nu^{(1)}(kr_>),
\end{equation}
in terms of the ultraspherical or Gegenbauer polynomial.  The stress
on the sphere is obtained by applying the appropriate differential
operator
corresponding to the stress tensor,
\begin{eqnarray}
T_{rr}&=&{i\over2}(\nabla_r\nabla_{r'}+\omega^2-
\bbox{\nabla}_\perp\cdot
\bbox{\nabla}_{\perp'}){\cal G}^0_{\omega}\nonumber\\
&\to&{i\over2}\left(r^{D-2}\partial_r r^{2-D}r^{\prime
D-2}\partial_{r'}
r^{\prime 2-D}+\omega^2-{n(n+D-2)\over r^2}\right){\cal G}^0_\omega.
\end{eqnarray}
Subtracting this from the previous result gives
\begin{eqnarray}
E_{\rm in}^{\rm TE}&=&-{1\over2\pi
a}{1\over\Gamma(D-1)}\sum_{n=1}^\infty
w(n,D)\int_0^\infty dx\,x\bigg\{{s_n'(x)\over s_n(x)}\nonumber\\
&&\quad\mbox{}+x^{3-D}\left[s_n'(x)e_n'(x)-\left(1+{n(n+D-2)\over
x^2}\right)
s_n(x)e_n(x)\right]\bigg\}.
\label{subte}
\end{eqnarray}

The question now is whether the continuation procedure described in
\cite{benmil} can be successfully applied here.  There, we first
made the integrals convergent by adding a suitable term to the
summand
which sums to zero for sufficiently small dimension.  Here this
suggests that in the above integral we replace
\begin{equation}
{s_n'(x)\over s_n(x)}={d\over dx}\ln x^{D/2-1}I_\nu(x)\to{d\over dx}
\ln\sqrt{2\pi x}I_\nu(x),
\end{equation}
for then the large $x$ behavior of this term
$1+(4\nu^2-1)/(8x^2)+\dots$.
The vacuum subtraction term cancels the leading term here, leaving
for
the leading term in the braces in (\ref{subte})
\begin{equation}
{(D-3)(D-2)\over4x^2}+O(x^{-4}).
\end{equation}
So, not surprisingly, the integral is still, in general,
logarithmically
divergent, {\em although for $D=3$ or $2$ it does converge.}

Therefore, it appears that we cannot meaningfully continue off the
integers.
So we are forced to retreat back to $D=3$.  There we have
\begin{equation}
E_{\rm in}^{D=3,{\rm TE}}=-{1\over2\pi a}\sum_{n=1}^\infty(2n+1)Q_n,
\end{equation}
where $Q_n$ is the convergent integral,
\begin{equation}
Q_n=\int_0^\infty dx\,x\left\{{d\over dx}\ln\sqrt{2\pi
x}I_\nu(x)+{\rm c.t.}
\right\}.
\end{equation}
If we use the uniform asymptotic expansions for the Bessel functions,
we easily find
\begin{equation}
Q_n\sim\nu^2\int_0^\infty dz\,z\left[{t^2\over8\nu
z}-{t^3(2t^2+3)\over4\nu^2
z}\right]\sim{\nu\pi\over16}+O(\nu^0),\quad (\nu\to\infty),
\end{equation}
where $t=(1+z^2)^{-1/2}$. The leading term here is precisely the
negative
of that found in the TM mode; that is, for electrodynamics, or
linearized
QCD, the interior modes contribute a quadratically divergent sum,
rather
than the cubically divergent one due to each mode.  Practical methods
of dealing with this divergent Casimir energy, which is relevant in
hadronic physics, were suggested in \cite{toward}.

\end{document}